\def\cm{cm$^{-1}$}

\def\bma{Ba\-Mn$_2$\-As$_{2}$}

\documentclass[prb,superscriptaddress,twocolumn,showpacs,amsmath,amssymb]{revtex4}
\usepackage[dvips]{graphicx}
\usepackage{graphicx}
\usepackage{color}

\begin{document}

\title{Optical properties of the iron-pnictide analog \bma}

\author{A. Antal}
\affiliation{1.~Physikalisches Institut, Universit\"at Stuttgart, Pfaffenwaldring 57, 70550 Stuttgart, Germany}
\affiliation{Institute of Physics, Budapest University of Technology and Economics
and Condensed Matter Research Group, Hungarian Academy of Sciences, 1521 Budapest, Hungary}
\author{T. Knoblauch}
\affiliation{1.~Physikalisches Institut, Universit\"at Stuttgart, Pfaffenwaldring 57, 70550 Stuttgart, Germany}
\author{Y. Singh}
\affiliation{I.~Physikalisches Institut,
Georg-August-Universit\"at G\"ottingen, 37077 G\"ottingen,
Germany}
\author{P. Gegenwart}
\affiliation{I.~Physikalisches Institut,
Georg-August-Universit\"at G\"ottingen, 37077 G\"ottingen,
Germany}
\author{D. Wu}
\email{dan.wu@pi1.physik.uni-stuttgart.de}
\affiliation{1.~Physikalisches Institut, Universit\"at Stuttgart, Pfaffenwaldring 57, 70550 Stuttgart, Germany}
\author{M. Dressel}
\affiliation{1.~Physikalisches Institut, Universit\"at Stuttgart, Pfaffenwaldring 57, 70550 Stuttgart, Germany}

\date{\today}

\begin{abstract}
We have investigated the infrared and Raman optical properties of \bma\ in the $ab$-plane and along the $c$-axis. The most prominent features in the infrared spectra are the $E_u$ and $A_{2u}$ phonon modes which show clear TO-LO splitting from the energy loss function analysis. All the phonon features we observed in infrared and Raman spectra are consistent with the calculated values. Compared to the iron-pnictide analog $A$Fe$_2$As$_2$, this compound is much more two-dimensional
in its electronic properties. For $E\parallel c$-axis, the overall infrared reflectivity is insulating like. Within the $ab$-plane the material exhibits a semiconducting behavior. An energy gap $2\Delta$=48~meV can be clearly identified below room temperature.
\end{abstract}

\maketitle
\section{Introduction}
The discovery of superconductivity in layered LaFeAsO$_{0.9}$F$_{0.1}$ four years ago has brought a new trend to synthesize and study new materials which are analogous to this compound.\cite{Hosono08,David10} By now, most of the interest is focussed on the materials as $AM_2X_2$ ``122''-type,  $A$Fe$X$ ``111''-type and Fe$X$ ``11''-type which have high superconducting transition temperature $T_c$ and can be obtained as high-quality single crystals. Only few studies have been reported on low $T_c$ pnictide compounds or those that do not show superconductivity at all.\cite{David10,Ronning08,Nath10,Singh09-1} Among those is \bma, that has the same tetragonal ThCr$_2$Si$_2$-type structure as $M$Fe$_2$As$_2$ at room temperature and undergoes no structural transition upon cooling.\cite{Singh09-1} The antiferromagnetic order appears at $T_N=625$~K; i.e.\ at much higher temperature than for the $M$Fe$_2$As$_2$ compounds.\cite{Singh09-2} While most of the $AM_2X_2$ parent compounds are metallic, \bma\ is a semiconductor with a small band gap $\Delta\approx$~27~meV according to  $dc$-resistivity measurement.\cite{Singh09-1} An $et~al.$ suggested that the difference in magnetic and electronic properties compared to those of $M$Fe$_2$As$_2$ is due to the strong Hund's coupling, the stability of the half-filled $d$-shell of the Mn$^{2+}$ ($d^5$ ion) and strong spin-dependent Mn-As hybridization.\cite{An09}

In this paper, we present mainly the infrared studies on \bma~single crystals with the
electric field polarized within the $ab$-plane and parallel to the $c$-axis. The sample with a size of $5\times5\times1~{\rm mm}^3$ was grown out of Sn fluxes. The magnetic, transport, and thermal properties of our material have been reported recently.\cite{Singh09-1,Singh09-2} For the infrared reflectance measurements, we cleaved the crystal within the $ab$-plane and finely polished the $ac$-plane in order to get shiny surfaces. The temperature dependent reflectivity spectra were measured in a wide frequency range from 40 to 37\,000~\cm\ using Fourier-transform infrared spectrometers ($40-15\,000$~\cm) and a variable-angle spectroscopic ellipsometer ($6000 - 37\,000$~\cm, restricted to room temperature). Raman spectra were measured using Jobin-Yvon T64000 spectrometer equipped with a microscope at room temperature. We used 514.5~nm Ar-laser excitation line with intensity below 1 mW. Spectra were measured in (a,a) and (b,b) polarizations. Several infrared and Raman phonon modes are assigned and compared to the calculated values.

\section{Experimental result and discussion}

\begin{figure}
 \centering
\includegraphics[width=0.9\columnwidth]{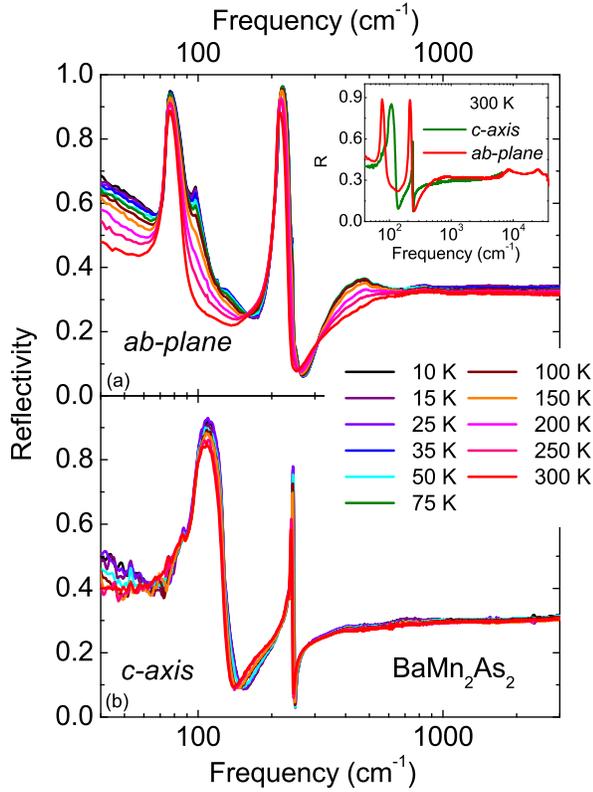}
 \caption{\label{fig:Fig1} (Color online)
Temperature dependent optical reflectivity of \bma\ measured in (a)~the $ab$-plane and (b) along the $c$-direction. The inset shows the room-temperature reflectivity in the whole measured range (40 - 37\,000 \cm). }
\end{figure}

In the inset of Fig.~\ref{fig:Fig1}, we plot the room temperature reflectivity for both $ab$-plane and $c$-axis polarization in the entire measured frequency range (40 to 37\,000~\cm). As approaching the zero frequency, $R_{ab}$ shows an upturn towards unity reflectance while $R_{c}$ tends to keep a constant value of approximately 40\%. The rather low overall reflectivity ($\sim$~30\%) indicates that \bma\ is a rather bad metal or a semiconductor. Besides the pronounced phonon vibrations in far-infrared region which we will discuss in detail later in this paper, two interband transition can be identified at approximately 7000~\cm\ (0.87~eV) and 26\,000~\cm\ (3.2~eV).\cite{Wu09,Li08} The temperature dependence of the reflectivity spectra is displayed in Fig.~\ref{fig:Fig1}(a) and (b). In the $ab$-plane, the reflectivity below 700~\cm\ increases with decreasing temperature, indicating a metallic behavior. This is consistent with the $dc$-resistivity measurement reported in Ref.~5. Only a very small temperature dependence is observed along the $c$-axis due to the less conducting property.

\subsection{Phonon vibrations}
\begin{figure}
 \centering
\includegraphics[width=1.0\columnwidth]{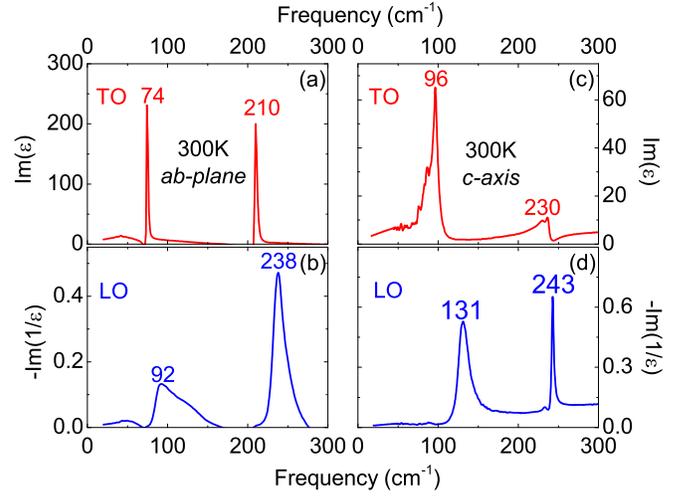}
 \caption{\label{fig:Fig5} (Color online)
Imaginary part of the dielectric function ${\rm Im}(\epsilon)$ and $-{\rm Im}(1/\epsilon)$ at 300~K for \bma~single crystal with (a,b) E$\|$ab and (c,d) E$\|$c. The frequencies of transverse and longitudinal optical modes are respectively identified as labeled in each panel. \cite{footnote1}}
\end{figure}

\begin{table}[t]
\caption{Comparison of the phonon mode frequencies (in \cm~) found in different \textit{AM$_2$X$_2$} compounds with the resonance frequencies of BaMn$_2$As$_2$. \label{table:Modes}}
\begin{center}
\begin{tabular}{|l|c|c|c|c|c|}
\hline\hline
Mode & CaFe$_2$As$_2$  &  SrFe$_2$As$_2$  & BaFe$_2$As$_2$  &  EuFe$_2$As$_2$ &  BaMn$_2$As$_2$ \\
\hline
A$_{1g}$ & 189  	& 182 	& 								& 				 & 174 \\
B$_{1g}$ & 211 	& 204  	& 209 				& 				 &  \\
E$_{g}$  &						& 114 	& 124 				& 				 &  \\
E$_{g}$  &						& 264 	& 264 				& 				 &     \\
\hline
A$_{2u}$ & 						& 					& 								& 				 & 96/131 \\
A$_{2u}$ & 						& 					& 								& 				 & 230/243 \\
E$_{u}$ & 						& 					& 94 	& 					 & 74/92 \\
E$_{u}$ & 						& 					& 253 , 254  	& 260 & 210/238  \\
\hline
Ref.    &12                     &13                 &14-16                          &8&this work\\
\hline
\hline
\end{tabular}
\end{center}
\end{table}
Due to the low reflection and weak absorption of \bma\ in the $ab$-polarization, a small fraction (about 5\%) of the far-mid infrared light was found to be transmitted through the sample during the quality-check measurement. This prevents us from a correct evaluation of conductivity spectra as commonly done by a Kramers-Kronig analysis.\cite{DresselGruner02} Nevertheless, all the pronounced phonon vibration modes show Lorentzian like shapes and thus are applicable for extracting the physics behind. According to the tetragonal ThCr$_2$Si$_2$-type structure (with space group I4/mmm) in which the $AM_2X_2$ usually crystallizes, a group theoretical analysis of the phonon modes in \bma~yields A$_{1g}$+B$_{1g}$+2E$_{g}$ Raman-active, 2A$_{2u}$+2E$_{u}$ infrared-active and A$_{2u}$+E$_{u}$ acoustic optical zone center phonons, where A/B and E modes correspond to an atomic motion perpendicular and parallel to MnAs planes, respectively.\cite{Chu08,Krou10} In our infrared spectra, however interestingly, we observe at least four vibration modes both in ab-plane and along the c-axis, as shown in Fig.~\ref{fig:Fig1}. Thus we extract the imaginary part of the dielectric function ${\rm Im}(\epsilon)$ and $-{\rm Im}(1/\epsilon)$ (known as the energy loss function) to identify the possible split transverse optical (TO) and longitudinal optical (LO) modes, respectively.\cite{Cardona} According to Fig.~\ref{fig:Fig5}, we can directly label the TO/LO energies to be 74/92 \cm~ and 210/238 \cm~for two E$_u$ modes, 96/131 \cm~and 230/243 \cm~ for two A$_{2u}$ modes.

\begin{figure}
 \centering
\includegraphics[width=0.8\columnwidth]{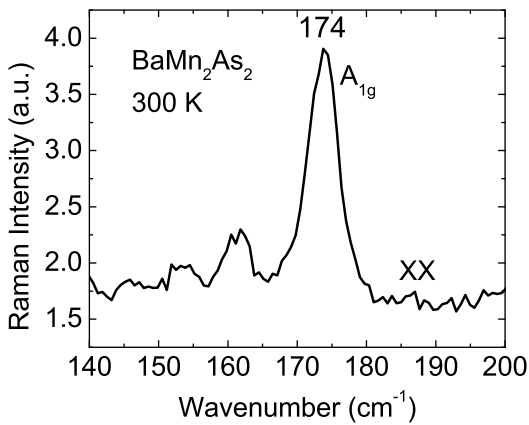}
 \caption{\label{fig:Fig7} (Color online)
XX-polarized Raman spectrum of \bma~at 300~K. The peak at 174 \cm~is assigned as A$_{1g}$ mode.}
\end{figure}

\begin{figure}
 \centering
\includegraphics[scale=0.6]{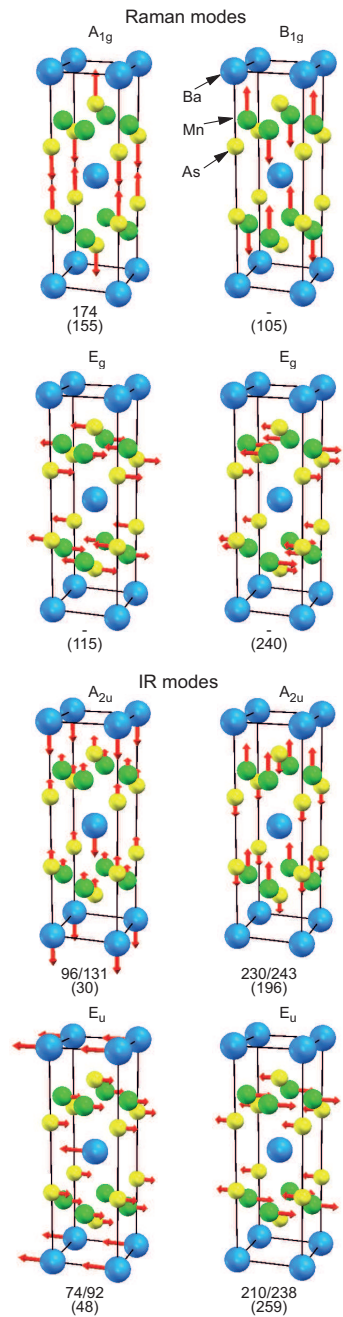}
 \caption{\label{fig:Fig6} (Color online)
Calculated atomic displacements and TO/LO frequencies in \cm~ for zone center phonons. Calculated frequencies are shown in brackets.}
\end{figure}

In Fig.~\ref{fig:Fig7} we show a room-temperature XX (A$_{1g}$ and B$_{1g}$ modes are allowed) Raman spectrum of \bma~ as obtained in (a,a) and (b,b) polarizations. A pronounced peak can be observed at 174 \cm. We refer this contribution to the A$_{1g}$ mode which is corresponding to the As-As displacement along c-axis (see also in Fig.~\ref{fig:Fig6}).

In Table.I a comparison of phonon modes frequencies found in \bma\ and several $A$Fe$_2$As$_2$ ($A$=Ca, Ba, Sr and Eu) compounds is shown. Due to the differences in lattice constant, atomic mass, and possible magnetic interactions (especially for EuFe$_2$As$_2$ and \bma~), the vibration frequencies for the same phonon mode are slightly different in these materials.

To compare with theoretical values, we also carried out phonon calculations (in the \emph{nonmagnetic} state (NM)) on the density-functional perturbation theory level within the plane-wave method with the generalized gradient approximation (GGA) of Perdew-Burke-Enzerhof.\cite{Baroni01,Perdew96} We used the quantum espresso package for the electronic structure calculation with the implemented ultrasoft plane-wave basis set.\cite{Giannozzi09} The band occupation was fixed and the literature lattice parameters has been used.\cite{Singh09} Only the atomic positions in the unit cell were optimized. The Brillouin zone is sampled in the \textbf{k} space with a uniform Monkhorst-Pack grid (12$\times$12$\times$12).\cite{Monkhorst76} In Fig.~\ref{fig:Fig6}, we display the calculated atomic transverse optical displacements and the measured TO/LO frequencies for all zone center phonons. Note that, in our calculations only a small LO-TO splitting of less than 1~\cm\ could be observed. The reason for such a small splitting and the discrepancy in the frequencies bases mainly on the nonmagnetic state revealing no band gap. An et al.\cite{An09} show by using density functional theory that the ground state of BaMn$_2$As$_2$ is antiferromagnetic (AFM) with a small bandgap of about 0.2 eV in the case of GGA calculations. Recent \textit{ab inito} calculations of the phonon spectrum in the ``122'' -family including the magnetic interaction lead to a better agreement of the phonon spectrum. \cite{Akteurk09,Mit09,Zbiri09} The difference in the mode frequencies between the NM and the AFM state can exceed 40~\cm.\cite{Kum11} It would be desirable to conduct further calculations including the AFM state which should lead to a small band gap and consequently a bigger LO-TO splitting.\cite{An09}

\begin{figure}
 \centering
\includegraphics[width=0.9\columnwidth]{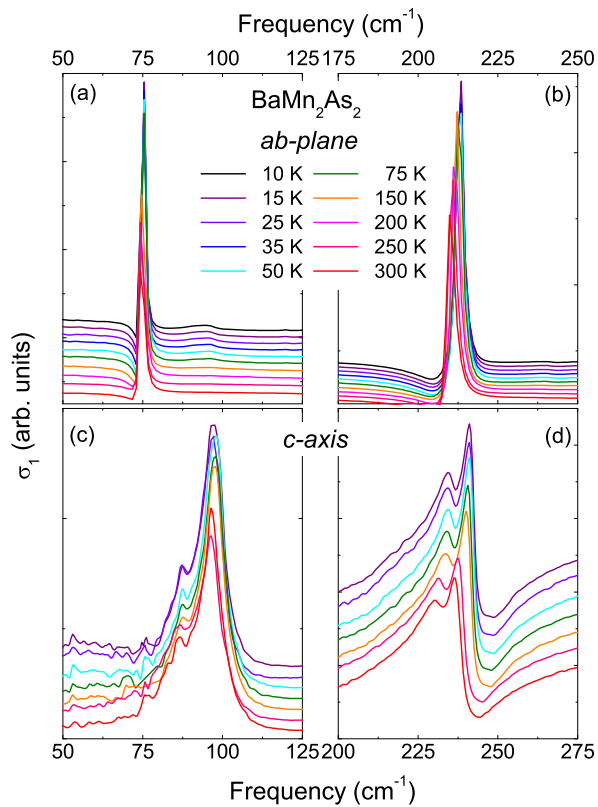}
 \caption{\label{fig:Fig4} (Color online)
Temperature development of conductivity in the region of phonon modes of \bma. }
\end{figure}

In Fig.~\ref{fig:Fig4}, no change in the phonon shape is observed for these modes due to the absence of structural transition upon cooling. But the E$_u$ mode with higher vibrational frequency is enhanced in intensity and continuously shift to higher frequencies. Recent neutron diffraction measurement on the thermal contraction revealed that
by cooling from 300~K to 10~K, $d_{\rm Mn-As}$ distance decrease from 2.566(2) to 2.558(2).\cite{Singh09-2}
The lattice distortion with reducing bond lengths enhances the electronic interactions between atoms and therefore the phonon vibrating energy increases -- consistent with the observed blue shift of this E$_u$ mode in our spectra.

\subsection{Energy gap}
\begin{figure}
 \centering
\includegraphics[width=0.9\columnwidth]{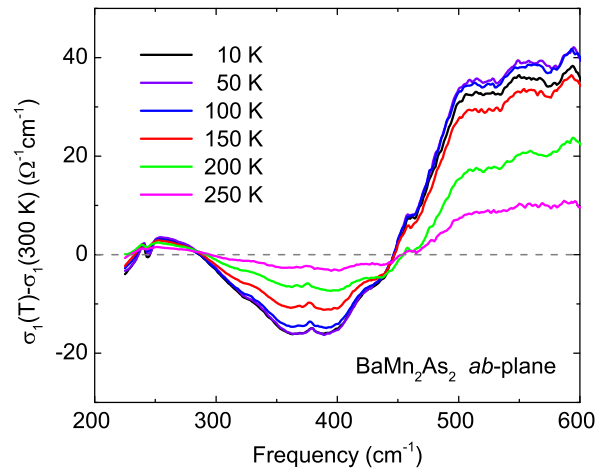}
 \caption{\label{fig:Fig2} (Color online)
 $\Delta\sigma_1(\omega,T)=\sigma_1(\omega,T) - \sigma_1(\omega,300~{\rm K})$ at selected temperatures. }
\end{figure}

There is an extra broad hump at the frequency around 400 \cm~ in the in-plane reflectivity spectra, as shown in Fig.1(a). According to the phonon calculations, we can exclude it to be a phonon mode. As the optical conductivity spectra of \bma~are phonon-rich ones, we will here analyze the optical data by looking at
the relative change of the conductivity $\Delta\sigma_1(\omega,T) = \sigma_1(\omega,T) - \sigma_1(\omega,T_0=300~{\rm K})$, as plotted Fig.~\ref{fig:Fig2}, to get rid of the strong phonon oscillations and see basically only the electronic background and consequently the spectral weight redistribution upon varying the temperature. As seen in Fig.~\ref{fig:Fig2}, around 300~\cm\  $\Delta\sigma_1 =\sigma_1(\omega,T)-\sigma_1(\omega,300~{\rm K})$ drops to negative values with decreasing temperature, then it increases at higher frequency and crosses zero at 440~\cm\, indicating that the spectral weight redistributes from low to high energies. Such a spectral weight rearrangement is the typical signature for an energy gap formation. From the minimum of $\Delta\sigma_1(10~{\rm K})$ we estimate the gap value 2$\Delta\sim390$~\cm\ (48~meV).\cite{Wu09,Hu08} Interestingly, Singh {\it et al.} recently performed a gap analysis on the $dc$-conductivity curve of this material.\cite{Singh09-1} There two linear ranges were found in the $\ln\{\sigma(T)\}$, and according to the expression $\ln\{\sigma\}=A-\Delta/k_B T$ two activation energies $\Delta=27$~meV and $\Delta=6.5$~meV were derived respectively. The gap value we yield from our optical experiments is in accord with the bigger one.\cite{footnote2}

\subsection{Anisotropy}

\begin{figure}[h]
 \centering
\includegraphics[width=0.9\columnwidth]{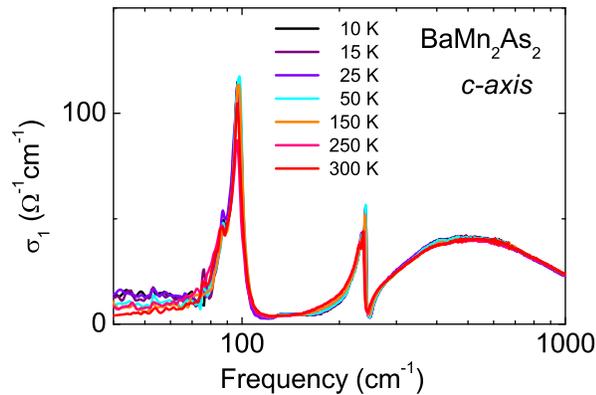}
 \caption{\label{fig:Fig3} (Color online)
$c$-axis conductivity of \bma\ as a function of frequency from 40 to 1000 \cm\ measured at selected temperatures.}
\end{figure}

The $c$-axis conductivity (Fig.~\ref{fig:Fig3}) varies only slightly with temperatures. When $\omega\rightarrow$~0~\cm, no Drude like behavior is observed and $\sigma_1(\omega\rightarrow~0)$ is quite low, indicating that the material is insulating along the $c$-axis.
No transfer of spectral weight is induced by the gap opening.
In contrast to iron-pnictides $M$Fe$_2$As$_2$, the present compound \bma\ reveals distinctively different electronic properties for the in-plane and out-of-plane directions, i.e.\ the material is more two-dimensional. In previous optical studies on BaFe$_2$As$_2$ by Wang and collaborators,\cite{Wang10-c}  the metallic responses were found for both $E\parallel ab$ and $E\parallel c$ with an anisotropy-factor about three.\cite{remark1} While two spin-density-wave gaps were revealed in the $ab$-plane, one of them also opens along the $c$-axis, giving evidence for a three-dimensional Fermi surface in the electronic structure.\cite{Wang10-c} Furthermore, the low carrier mobility in \bma\ infers more localized charge carriers, in the contrast to the itinerant nature of the electronic properties of $M$Fe$_2$As$_2$.
The reason is the antiferromagnetic arrangement of the Mn$^{2+}$ spins both in the $ab$-plane and along the $c$-axis (G-type antiferromagnetic order), where at low temperature the electron hopping between nearest neighbor sites with opposite spin is suppressed and a strong scattering is induced by the spin disorder at high temperature.\cite{An09}

\section{Summary }
In summary, the optical properties of iron-pnictide analog \bma\ have been studied in detail using light polarized in the $ab$-plane and along the $c$-axis. Within the $ab$-plane, the material behaves like a poor metal or semiconductor. An energy gap can be defined at $2\Delta$=48~meV. The $c$-axis spectra show a typical insulating behavior overall and no energy gap is found. This indicates that the \bma\ is a two-dimensional electron system rather than a quasi three-dimensional, as the Fe-pnictides are considered to be. The $E_u$ and $A_{2u}$ phonon modes have clear TO/LO splitting with frequencies differences of 10-40 \cm. A systematic comparison between experimental results and theoretical calculations are present for all the infrared and Raman modes.

\begin{acknowledgments}
We thank B. Gorshunov and V. I. Torgashev for very helpful discussions. We appreciate N. Drichko at Johns Hopkins University, Baltimore MD, for performing the Raman measurement. The
contributions of J. Braun and D. Rausch to the experiments are
appreciated. D.W acknowledges her fellowship by the Alexander von
Humboldt Foundation. A.A. acknowledges the support of the German Academic Exchange Service (DAAD) and also Hungarian National Research Fund OTKA NN76727. T.K. acknowledges his scholarship by the Carl-Zeiss-Stiftung. Part of the work was funded by the Deutsche Forschungsgemeinschaft (DFG).
\end{acknowledgments}


\begin{thebibliography}{99}
\bibitem{footnote1} In Fig.~\ref{fig:Fig5}(a) and (b), there are ranges (for example, 170-210 \cm) where ${\rm Im}(\epsilon)$ shows negative values. This unphysical phenomenon is caused by the transparent property of sample as we discussed at the beginning of II-A section.
\bibitem{remark1}In the case of EuFe$_2$As$_2$ the anisotropy was about a factor of 8 (Ref.~\protect\onlinecite{Wu09}).
\end{thebibliography}
\end{document}